# Risk time splitting for improved estimation of screening programs' effect on later mortality

Harald Weedon-Fekjaer[1], Elsebeth Lynge[2], Niels Keiding[3†]

1) Oslo Centre for biostatistics and epidemiology, Research Support Services, Oslo University Hospital, Oslo, Norway

2) Centre for Epidemiology and Screening, Department of Public Health, University of Copenhagen, Copenhagen, Denmark

3) Section of Biostatistics, Department of Public Health, University of Copenhagen, Denmark

†) Niels Keiding died 3 March 2022. Niels suggested this paper reading the analysis introduced in Weedon-Fekjær et al.'s 2014 BMJ paper, participated as planned co-author in multi-day meetings, set the new method into the context of other methods, and made key contributions to the writing of the paper.

Correspondence:     Harald Weedon-Fekjær

Oslo Centre for biostatistics and epidemiology, Research Support Services, Oslo University Hospital, Oslo, Norway

E-mail:          <harald.weedon-fekjar@medisin.uio.no>

## *Abstract*

There is a great need for evaluating screening programs, but analysing data from population screening is often complicated by a delayed screening effect. In cancer screening, only new, not yet clinically diagnosed cases, might benefit from screening through earlier treatment. Hence, mortality data following screening should be analysed based on refined mortality, separating cases based on diagnosis before and after first screening invitation. Historically, refined mortality has been implemented by selecting comparison groups from the available data to disentangle the causal effect. While giving valid estimates, the ignorance of large parts of the available data has limited study precision. In BMJ 2014, Weedon-Fekjær et al. used a new estimation approach applying all the available Norwegian mammography screening data. The estimation uses historic pre-screening data on time from clinical diagnosis to death estimating the proportion of post-screening mortality which is expected to be based on cases incident before first screening invitation, in the absence of a screening effect. Utilizing this expected proportion of post-screening incident cases, Poisson regression offsets are added to align the expected number of cases. The screening effect is then estimated adjusting for relevant covariables. While the method increases study precision, it has not been easily available and widely adopted. We here explain the method in detail, add maximum likelihood estimation, and lay the foundation for widespread use. Applying the method on Norwegian and Danish data, bootstrap confidence intervals are considerably narrower than intervals seen using other refined mortality methods, especially for the gradually introduced Norwegian program.

## 1. Introduction

Early detection of disease through screening might save lives and increase life quality by enabling timely treatment. Lately, screening has been suggested for a growing number of diseases from cancers and cardiovascular disease to mental health problems and early life development[1]. Statistical analysis is, however, not straightforward as cases with potential screening effect are usually mixed with old pre-screening diagnosed cases with no potential screening effect.

For cancers, screening has been implemented for several cancer types based on the observed better survival of early-stage cases in clinical datasets. The hope is that earlier detection improves the prognosis through earlier treatment, reducing population mortality for the given cancer type. The causal link between earlier treatment and improved prognosis is, however, not evident and should be carefully evaluated. The gold standard of medical interventions is randomized controlled trials, but randomized trials are not always feasible and cannot be used to evaluate already ongoing population-wide service screening. One example is breast cancer mammography screening, which today is implemented in most developed countries based on the reduced breast cancer mortality observed in old randomized controlled trials[2,3]. By time, breast cancer treatment has changed considerably since the old randomized controlled trials[4]. It is therefore pertinent to monitor the outcome of service screening, as mammography screening is only justified if it has a beneficial effect on population wide breast cancer mortality on the top of newly implemented treatment regimes[5,6]. It is often possible to measure the effect of introducing service screening, as most programs are gradually implemented due to the logistic challenges of training competent staff. Several studies have taken advantage of the gradual implementation of the Norwegian, Danish, Swedish and Finnish service screening programs, using observations from the not yet invited women as the control group when estimating the effect of the screening program on breast cancer mortality[7-10]. The analysis design is, however, not straightforward.

Some studies of public screening programs have looked at general cause-specific mortality trends[11], but mixing cases incident before and after screening introduction strongly dilutes the estimates of the causal screening effect, possibly hiding real screening effects[12,13] (Figure 1). Others have implemented ad-hoc adjustments for screening related lead time[9,14]. Most modern studies use methods based on refined mortality[15], splitting cases diagnosed before and after first screening invitation. Typically these studies carefully selected matched comparison groups pre- and post-screening introduction[7,8,15-17]. While the refined mortality studies are methodically valid with consistent estimators, their neglect of part of the available data often results in limited statistical precision with wide confidence intervals[8,17,18].

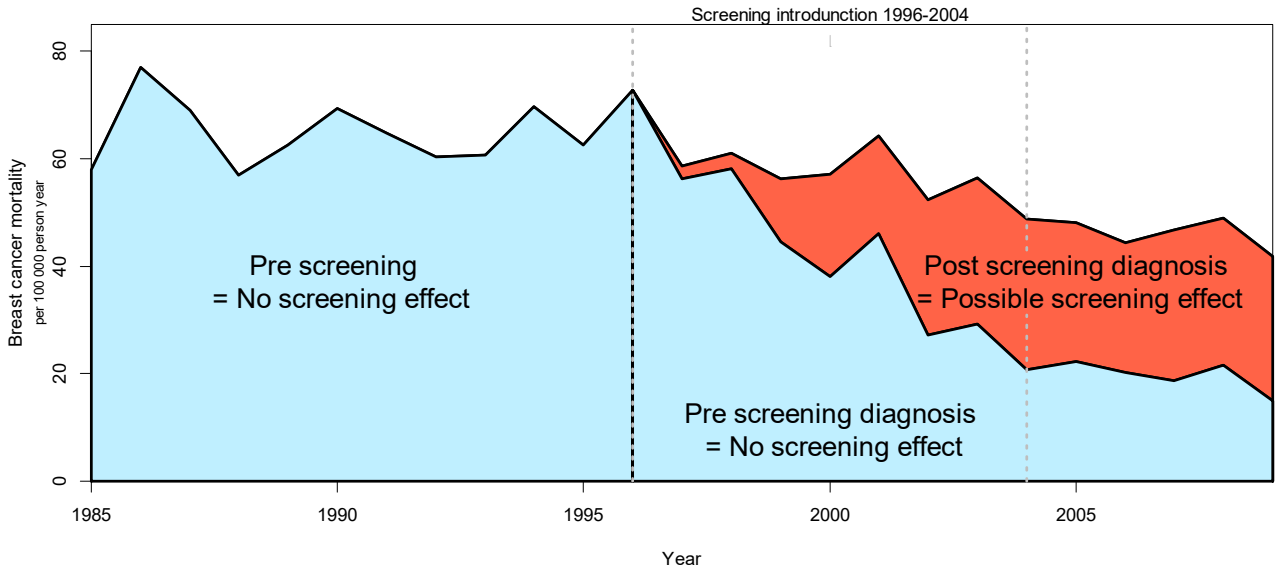

Figure 1:    Norwegian breast cancer mortality around screening introduction split by cases with diagnosis pre and post first screening invitation.

A new estimating approach was developed and applied by Weedon-Fekjær et. al. BMJ 2014[19], greatly improving the utilization of the data without adding additional strong modelling assumptions. The technical details of this method were mostly hidden away in a web appendix, complicating new applications of the suggested method. Hence, the purpose of this paper is to provide a more in depth and complete presentation of this new approach for a biostatistical audience, making the method widely accessible. We begin by presenting a simplified, more intuitive version (Method I), before we go on to the full method applied in BMJ 2014[19] (Method II). We then deduce a corresponding full maximum likelihood approach (Method III), before we illustrate and compare the approaches using Danish and Norwegian breast cancer data. Last example code, showing how to implement the recommended method in both the R statistical software package and the Python programming language, is given.

## 2.  *Earlier studies estimating cancer screening effects – the "classical" analysis approaches*

Screening programs aim to improve survival by enabling more effective treatment through earlier detection at screening. Hence, screening is only expected to influence mortality of cases not yet diagnosed at the time of screening introduction. When estimating screening programs' efficiency in reducing cancer mortality, cause specific population mortality of post-screening incident cases is usually selected as outcome measurement to get reasonable statistical power[7,8,15-17]. The screening effect is commonly estimated as the ratio between observed and expected cancer type specific mortality among new incident cases. Usually this ratio is assumed to be constant by time, fitting well with observations from breast cancer screening trials[2]. Competing risks as death to other causes or emigration are typically treated as non-informative censoring.

As most women dying of breast cancer have been diagnosed several years ago, screening effects are expected to come very gradually into the population, complicating the analysis[12]. Some studies of breast cancer screening programs have looked directly at overall time trends in cancer mortality[11], but these studies are highly diluted and underpowered by cases with no potential screening effect[13]. Hence, breast cancer screening programs should never be evaluated without splitting cases diagnosed pre- and post-screening introduction[13].

Other studies have compared breast cancer mortality between screening program attendees and invited non-attendees[20]. While the attendees versus non-attendees comparison might easily be restricted to new incident cases, selective attendance can substantially bias the estimated screening effects. Some of these studies adjust for historic differences in breast cancer mortality across attendance, but the selection effects might be very different across different decades and populations.

The breast cancer mortality varies substantially over time for different age groups, time periods and counties, due to both variations in disease risk factors and new treatments[4,21]. One example is tamoxifen treatment, which substantially improved breast cancer survival when it was introduced around the turn of the century[22]. Hence, analysis of non-randomized breast cancer screening programs' effects should at least always adjust for variations in calendar time, attained age and birth cohort[23,24].

Optimally, analysis should compare invited versus non invited women, using *refined mortality*[25] carefully splitting cases diagnosed before and after screening introduction. The challenge with refined mortality analysis is to ensure that the groups are comparable across age groups, time periods and counties, as the number of cases excluded due to pre-screening diagnosis varies by age and time since screening introduction. For population screening programs, the invited and non-invited groups have often been compared by mimicking randomized screening trials, carefully selecting comparison groups from the available data[7,8,17,26]. In these studies, rates of refined mortality in screening counties are compared to mortality based on new incidence cases in non-screened control counties given an artificially set cut-off date for new incident cases matching the screened group. Then the mortality rates are typically adjusted for the time trends seen in non-screening counties, often using a ratio of ratios. This recent vs. historic comparison, across populations with and without an intervention, is well known in econometrics as difference in differences analysis[27,28]. Implemented carefully, this double ratio yields a consistent estimate, but the selection of comparison groups leave much of the available data unused, limiting the studies' statistical power.

### *3. Estimating screening effect by comparing predicted standardized mortality (Method I)*

While the earlier refined mortality studies only used selected part of the available data [7,8,17], Weedon-Fekjær at. al.'s analysis of BreastScreen Norway introduced a new method applying all the available data[19]. Before presenting the full Weedon-Fekjær et. al. refined mortality regression approach in detail, we here first present a simpler more intuitive approach that roughly approximates the more precise estimating method.

Evaluating screening, the observed number of the given cause of death based on post-screening diagnosis (with potential screening effect) should be compared with the corresponding expected number of cases given no screening effect, adjusted for relevant covariables. As mortality is known to vary by age, birth cohort and time period, we use Age-Period-Cohort models assuming constant rate of the relevant cause of death within each age-period-cohort combination, modelled by a set of age, period and cohort parameters[24]. Applying Poisson regression on cause specific mortality by age, cohort, period and region, the expected cause specific mortality in the post-screening period can be estimated. The derived expected mortality might be compared with the observed mortality, but the estimated screening effect will be highly diluted due to pre-screening diagnosed cases with no potential screening effect. However, using historic retrospective data on the time since diagnosis for the given cause of death, a three-step approach can be used to study only cases incident post first screening invitation with a potential screening effect:

I.  Based on data from not yet screened cohorts, the expected number of the relevant cause of death in the absence of screening effects is estimated for the post-screening invitation group, using an age-period-cohort-region Poisson regression model with relevant covariables (Figure 2).

II.  Applying retrospective times since diagnosis for the given cause of death among non-screening invited women, the probability of a case being based on a pre-screening diagnosis in the absence of screening effects is estimated. Using these estimates, the expected proportion of cases with pre-screening invitation diagnosis in the post-screening mortality group is estimated for each age, period, cohort and region combination. Subtracting the expected proportion of cases with pre-screening invitation diagnosis, the expected number of cases for the relevant cause of death based on diagnosis post-screening invitation in the absence of a screening effects is calculated.

III.  Comparing the observed number of the relevant cause of death based on post-screening diagnoses with the corresponding expected number of cases given no screening effects, the average effect of the screening program could be estimated as a rate ratio. Splitting the post-screening data into sub datasets, the screening effect by time since screening introduction, or calendar year, might be estimated. However, most real-life studies of screening programs do not have enough statistical power for sub analysis.

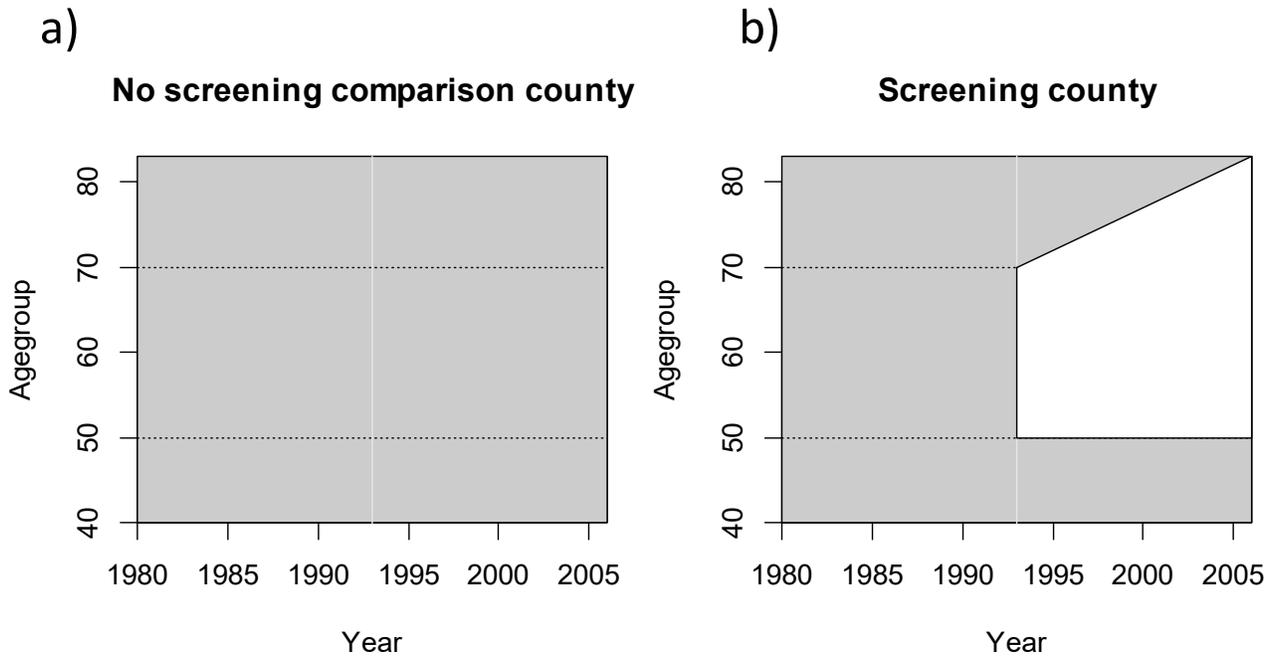

Figure 2: Example of non-screened (grey) and screened (white) populations by age and calendar time for two different counties, used as comparison groups in the simplified standardized mortality method for estimating screening effects (I); Based on the non-screening data (grey), the expected number of cases in the absence of screening effects is estimated, before the observed versus expected ratio is calculated for the screened population. When using refined mortality for the screening data, the expected number of cases in the screening data is adjusted by the expected proportion of cases with a pre-screening invitation diagnosis in the absence of a screening effect.

For the analysis, we typically model the observed rates of cause specific mortality by Poisson regression using smoothed splines on age, period and cohort to limit the number of free model parameters and stabilize the estimation[29]:

$$M_{c,p,r} = \exp( A(a) + P(p) + C(c) + R_r )  \qquad (1)$$

where $M_{c,p,r}$ is the expected cause specific mortality for cohort $c$ in period $p$ for age $a = p - c$, $R_r$ is the log baseline cause specific mortality for region $r$, and $A(a), P(p)$ and $C(c)$ are natural smoothing splines modelling the gradual changes in mortality by age, calendar time and birth cohort.

Applying Poisson regression on the pre-screening data, we obtain estimated values of the age effects $A(a)$, the period effects $P(p)$, the cohort effects $C(c)$, and the regional level $R_r$, denoted $\hat{A}(a), \hat{P}(p),$ $\hat{C}(c)$ and $\hat{R}_r$, respectively. The full age-period-cohort model is over-parameterized, leading to an identifiability problem. However, when analysis is restricted to the observed age groups, periods, and cohorts, the specific allocation of the estimated effects among the age, period, and cohort parameters becomes inconsequential[24,30]. Using the estimated model parameters, the expected number post-screening deaths in the absence of a screening effect could now be estimated by:

$$\hat{M}_{c,p,r} = \exp\left( \hat{A}(a) + \hat{P}(p) + \hat{C}(c) + \hat{R}_r \right) \tag{2}$$

As we only want to use post-screening incident cases in our screening effect estimate, the expected mortality must be corrected for the expected proportion of cases being incident pre-screening. Defining $Age_{\text{diagnosis}}$ as the age at diagnosis for women later dying of the relevant cause of death under study at age $Age_{death}$, we define the time from diagnosis to death as $\tau = Age_{death} - Age_{\text{diagnosis}}$. In practice, we are interested in the proportion of cases with a $\tau$ greater than the elapsed time from screening initiation. Setting $\rho_{\delta}$ as the expected proportion of deaths with a diagnosis more than $\delta$ months ago in the absence screening effects, the proportion $\rho_{\delta} = P(\tau > \delta)$, can be directly estimated based on pre-screening data by:

$$\hat{\rho}_{\delta} = \frac{\text{number of cause specific deaths with diagnosis more than } \delta \text{ months ago}}{\text{total number of cause specific deaths}} \tag{3}$$

Combining $\hat{\rho}_{\delta}$ and $\hat{M}_{c,p,r}$, we can now estimate the expected mortality based on diagnosis post-screening invitation in absence of screening effects, $\widehat{MPOST}$, by:

$$\widehat{MPOST} = \sum_{c,p,r \in R_{screened}} \hat{M}_{c,p,r} \left(1 - \hat{\rho}_{\text{tsscr}_{c,p,r}}\right) \tag{4}$$

where $tsscr_{c,p,r}$ is the mean time since screening introduction for cohort $c$ in period $p$ within region $r$, and $R_{screened}$ is the screening invited combinations of $c$, $p$ and $r$. Using the observed mortality rate based on diagnosis post screening, $MPOST$, the screening effect, $ScrEff$, can be estimated as rate ratio:

$$\widehat{ScrEff} = \frac{MPOST}{\widehat{MPOST}} \tag{5}$$

The estimate might be seen as a standardized mortality Ratio [SMR] based on the coefficients from the Poisson regression analysis and $\hat{\rho}$[31,32]. Associated confidence intervals can be deduced using basic nonparametric statistical bootstrap, resampling the data behind both $\hat{M}$, $MPOST$ and $\hat{\rho}$ assuming Poisson ($\hat{M}$, $MPOST$) and Binomial distributed ($\hat{\rho}$) data. In practice, $\rho_{\delta}$ will vary somewhat with age, so $\rho_{\delta,ai}$ could be estimated separately for different age intervals $ai$, with $\widehat{MPOST}$ calculated as the sum across all age groups. In our calculations, we used 10-year age groups for $\rho_{\delta,ai}$. To increase precision, we applied one-month intervals for the lag from diagnosis to death when calculating $\rho_{\delta,ai}$.

The idea behind this approach have similarities to negative Outcome Control [33,34], where the observed effects are checked using comparison groups that are not believed to be affected by the exposure under study. Here, the non-invited women can be seen as negative controls with no assumed screening effect, but similar effects of age, period, cohort and region.

While screening might work differently across age groups, time periods, and times since screening initiation, our suggested approach only gives one weighted average across these different subgroups. Subgroup analysis might be added by defining different screening effect parameters for each subgroup of interest. Data is, however, limited in most settings, and for at least breast cancer randomized trials data indicate limited variations in the relative screening effect by time since screening introduction[2]. Typically, one might add some exploratory analysis of selected age groups and calendar time periods in the lack of statistical power for more detailed analysis.

## 4. *Refined mortality regression analysis of screening effects (Method II)*

While the simplified method above (method I) gives a valid estimate, rate ratios are not very stable, and the estimation could be further improved by a more coherent estimation of pre- and post-screening incident mortality. As for the simple method, any efficient estimator of screening effects should take into account whether the given cause specific deaths are based on diagnoses pre- or post-screening introduction to avoid dilution of the screening effect estimate. However, when performing Poisson regression on data split by time of diagnosis, person years post screening invitation could give rise to deaths based on both cases diagnosed pre- or post- screening invitation. Soon after initial screening, most cause specific deaths would be based on diagnosis made pre-screening, while by time the expected proportion of post-screening diagnosed cases is expected to increase. In most datasets, this gradually increasing proportion of post-screening incident cases, $\rho_{\delta,a}$, by time, $\delta$, is supported by quite a bit of pre-screening data for each age interval, $a$, making it possible to estimate reliably using equation ( 3 ).

Conditioning on a given set of $\hat{\rho}$ estimates, the screening effect might be estimated using Poisson regression analysis in standard statistical software with only a slightly customized input. Since we for this method also use the post-screening cases based on pre-screening diagnosis, we gain some extra information regarding the underlying cause specific mortality trends. The rates of cases diagnosed pre- and post-screening introduction is, however, neither directly comparable to each other or directly comparable to the pre-screening rate. Hence, the post-screening rates split by time of diagnosis must be adjusted before comparing them to the pre-screening rate.

In epidemiology, standardization of rates aims to adjust rates making them comparable to expected rates for a commonly defined reference population[35]. Typically, rates are adjusted for different age structures when compared across different populations. As with age standardization routines, we here also adjust the rates using the estimated probability of belonging to the given group. Hence, the expected rates in the Poisson regression model are multiplied with the estimated likely proportion of cause specific deaths with a diagnosis more, or less, than the given time from screening introduction, $\delta$, in the absence screening effects, $\hat{\rho}_\delta$. Using $\hat{\rho}_\delta$ from equation 3, we adjusts the expected rate of cases incident pre-screening by $\hat{\rho}_\delta$, and the post-screening incident cases by 1 - $\hat{\rho}_\delta$. As $\hat{\rho}_\delta$ might vary by age, this is typical done separately for different age intervals $ai$, each with its own estimated $\hat{\rho}_{\delta,ai}$. Using standard Poisson regression software, the adjustment could be added as Poisson regression offsets. Assuming a uniform screening effect among post-screening incidence cases across age, cohort and time since screening, the screening effect, $ScrEff$, can now be estimated by adding an indicator equalling one only for the post-screening incidence cases. Our overall model for incident based mortality now becomes the combination of three different rates; Pre-screening cause specific mortality, **MIPreScr**, post-screening cause specific mortality based on old cases incident pre-screening, **MIPostOld**, and post-screening cause specific mortality based on new cases incident post-screening, **MIPostNew**, given as:

$$MIPreScr_{c,p,r} \ = \ \exp( \ A(a) + P(p) + C(c) + R_r \ ) \qquad (\,6\,)$$

for all combinations of *c*, *p* and *r* without a screening history, and:

$$MIPostOld_{c,p,r,\delta,ai} = \ \exp( \ A(a) + P(p) + C(c) + R_r \ ) \times \hat{\rho}_{\delta,ai} \qquad (\,7\,)$$

and

$$MIPostNew_{c,p,r,\delta,ai} \ = \ \exp( \ A(a) + P(p) + C(c) + R_r + ScrEff \ ) \ \times \ \left( 1 - \hat{\rho}_{\delta,ai} \right) \qquad (\,8\,)$$

for all combinations of *c*, *p* and *r* with a screening history.

Where:

- $A(a), P(p)$ and $C(c)$ representing the smoothed effect of age, $a$, calendar time, $p$, and birth cohort, $c$, on cause specific mortality by for example natural smoothing splines, where age $a = p - c$

- $R_r$ denotes the baseline cause specific mortality for region $r$

- $ScrEff$ is the effect of the screening program on new post-screening incident cause specific mortality

- and $\hat{\rho}_{\delta,ai}$ the estimated expected proportion of cases with pre-screening diagnosis in the lack of a screening effect with a given time of $\delta$ since first screening invitation for age interval $ai$. Here $\hat{\rho}_{\delta,ai}$ is estimated by equation ( 3 ), with $\delta$ defined by the observed screening history for cohort $c$ at period $p$ in region $r$.

**Implementing the refined mortality regression analysis of screening effects:**

In practice, refined mortality regression analysis could be implemented by:

a) Gather cause specific mortality data by calendar period, birth cohort and other potentially relevant co-variables as region split by cases pre-screening, cases post-screening based on diagnoses pre-screening, and cases post-screening based on diagnoses post screening. As either age, period and cohort are redundant due to their linear relationship, data might either be sampled by age-period or period-cohort, typically depending on how the screening intervention is introduced into the population. Note that any case found on initial screening should be counted as a post screening invitation case.

b) Add number of person years at risk for each combination.

c) Add an indicator for potential screening effect, equalling one only for the data concerning post-screening cases based on diagnoses post screening, that is multiplied with a screening effect parameter.

d) Add time since first screening invitation for the post-screening introduction data

e) Gather historic data on time from diagnosis to death among persons dying from the relevant cause specific mortality. Data should be collected by time of death, not time of diagnosis, to avoid biases by potential long unobserved lags from diagnosis to death. Optimally chose data close to the screening introduction, to include any effects of new treatments and other changes by time.

f) Using the historic pre-screening data on time from diagnosis to death, calculate $\hat{\rho}_{\delta,ai}$, the expected proportion of the cause specific mortality that is based on diagnoses more than $\delta$ months ago in the absence of screening for each age group $ai$.

g) Include $\hat{\boldsymbol{\rho}}$ to our cause specific mortality data (from step a-d), for use as offsets in the regression model adjusting the expected number of cases:

- For the pre-screening introduction data set the offset to 1.

- For the post-screening introduction data with cases incident pre-screening set the offset to $\hat{\rho}_{\delta,ai}$.

- For the post-screening introduction data with cases incident post-screening set the offset to $1-\hat{\rho}_{\delta,ai}$.

with the time since screening introduction, $\delta$, calculated individually for each combination of age, period, cohort and region, for the relevant age group $ai$ (Figure 3).

h) Find the estimated log screening effect on cause specific mortality using Poisson regression, with the given covariates and offsets. The estimated screening effect can now be deduced as the exponential of the estimated log screening effect, with confidence intervals calculated by basic non-parametric statistical bootstrap replications, resampling all the applied data sources assuming Poisson and Binomial distributions.

Example of R and Python programming code for the estimation of screening effects, using the refined mortality regression analysis approach (Method II) is given in the appendix.

### a) Data formatted for refined mortality analysis

| Year | Age | Screening group | Population | Cases | Time since screening invitation | Relative expected incidence | Screening indicator |
|---|---|---|---|---|---|---|---|
| (co-variable) | (co-variable) | (internal variable) | (co-variable) | (co-variable) | (internal variable) | (regression offset) | (main co-variable) |
| 1988 | 55 | No screening | 2297 | 3 | | 1.00 | 0 |
| 1989 | 55 | No screening | 2302 | 3 | | 1.00 | 0 |
| 1990 | 55 | No screening | 2366 | 2 | | 1.00 | 0 |
| 1991 | 55 | No screening | 2046 | 0 | | 1.00 | 0 |
| 1991 | 55 | Post | 201 | 0 | 0.2 | 0.08 | 1 |
| 1991 | 55 | Pre | 201 | 0 | 0.2 | 0.92 | 0 |
| 1992 | 55 | No screening | 1211 | 1 | | 1.00 | 0 |
| 1992 | 55 | Post | 1006 | 0 | 0.5 | 0.11 | 1 |
| 1992 | 55 | Pre | 1006 | 1 | 0.5 | 0.89 | 0 |
| 1993 | 55 | No screening | 420 | 0 | | 1.00 | 0 |
| 1993 | 55 | Post | 1909 | 0 | 1.1 | 0.17 | 1 |
| 1993 | 55 | Pre | 1909 | 3 | 1.1 | 0.83 | 0 |

### b) Expected proportion new incident cases by time since screening introduction

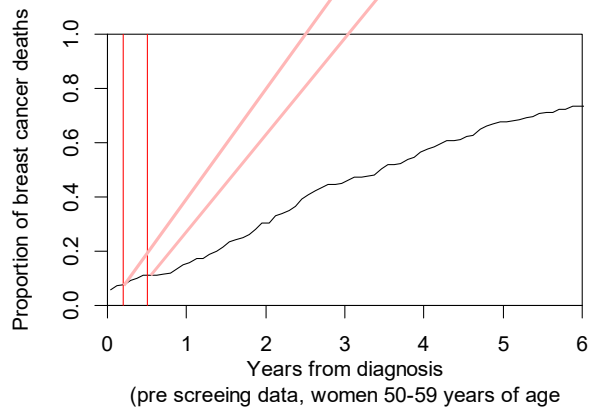

Figure 3: Exemplar of input data for the regression analysis of screening effects based on refined mortality (method II). In the regression approach, historic data on the lag from breast cancer diagnosis to breast cancer death (b) is used, as offsets to align the refined mortality analysis data (a) to the pre-screening data after splitting cases diagnosed pre and post first screening invitation (Danish cancer registry data with some added random noise to mask personal identifiable data).

## *5.* *Maximum likelihood estimation of screening effects (Method III)*

While giving consistent estimates, the refined mortality regression analysis approach (method II) might be improved. Hence, we here derive an estimator based on the commonly acknowledged maximum likelihood approach[36-38]. We base the new maximum likelihood estimator on the same data, principles and set of assumptions as the previous presented regression analysis approach (method II).

As with the refined mortality regression analysis of screening effects, population mortality data is first split by calendar period, birth cohort, region, and whether post-screening cases were diagnosed pre or post first screening invitation. The overall likelihood becomes a combination of the likelihood for the population mortality data, and the likelihood for the historic data regarding time from diagnosis to death among not yet screened cohorts, using a common set of parameters describing the lag from diagnosis to cause specific death in the absence of screening interventions:

$$P\big(\boldsymbol{data}_{\text{mortality}}, \boldsymbol{data}_{\text{mortalityLag}} \mid \boldsymbol{\theta}_{\text{mortality}}, \boldsymbol{\theta}_{\text{mortalityLag}}, \theta_{scren\_effect}\big) \qquad (\,9\,)$$

$$= P\big(\boldsymbol{data}_{\text{mortality}} \mid \boldsymbol{\theta}_{\text{mortalit}} , \boldsymbol{\theta}_{\text{mortalityLag}} , \boldsymbol{\theta}_{scren\_effect}\big)$$
$$\cdot \ P\big(\boldsymbol{data}_{\text{mortalityLag}} \mid \boldsymbol{\theta}_{\text{mortalityLag}}\big)$$

Where:

- $\boldsymbol{data}_{\text{mortality}}$ is the period-cohort population mortality data by region.

- $\boldsymbol{data}_{\text{mortalityLag}}$ is the data on time from diagnosis to cause specific death pre-screening introduction.

- $\boldsymbol{\theta}_{\text{mortality}}$ is the vector of age, period, cohort, region parameters describing the Age-Period-Cohort-Region variations in the cause specific mortality.

- $\boldsymbol{\theta}_{\text{mortalityLag}}$ is the vector of parameters describing the lag data from diagnosis to cause specific death in the absence of screening effects.

- and $\theta_{scren\_effect}$ is the parameter for the (potential) screening effect.

**Likelihood for the population mortality data**

The likelihood for the population mortality data is a product of the likelihood of the pre-screening population data, post-screening population data counting only cases incident pre-screening, and post-screening population data counting only cases incident post-screening:

$$P\left(\boldsymbol{data}_{mortalit} \mid \boldsymbol{\theta}_{mortalit}, \boldsymbol{\theta}_{mortalityLag}, \boldsymbol{\theta}_{scren\_effect}\right) = \qquad (10)$$

$$P\left(\boldsymbol{data\_noscr}_{mortality} \mid \boldsymbol{\theta}_{mortality}\right) \cdot$$

$$P\left(\boldsymbol{data\_pre}_{mortality} \mid \boldsymbol{\theta}_{mortality}, \boldsymbol{\theta}_{mortalityLag}\right) \cdot$$

$$P\left(\boldsymbol{data\_post}_{mortality} \mid \boldsymbol{\theta}_{mortality}, \boldsymbol{\theta}_{mortalityLag}, \boldsymbol{\theta}_{scren_{effect}}\right)$$

Where:

- $\boldsymbol{data\_noscr}_{mortality}$ is the pre-screening population data

- $\boldsymbol{data\_pre}_{mortality}$ is the post-screening population data counting only cases incident pre-screening

- $\boldsymbol{data\_post}_{mortality}$ is the post-screening population data only cases incident post-screening

Splitting the mortality data by period, cohort, region and diagnosis pre vs. post first screening invitation, we can organize it as a vector of data, $\boldsymbol{data}_{mortality} = \{X_{p,c,r,s}\}$, for all relevant combinations of period $p$, cohort $c$, region $r$ and screening group $s$ (pre-screening, post-screening with pre-screening diagnosis or post-screening with post-screening diagnosis). Assuming a piecewise constant hazard rate within each combination, the number of cases in each cell follows a Poisson distribution[39]. With the expected number of cases given as $\mu_{p,c,r,s}$, this gives a likelihood of:

$$P\left(\boldsymbol{data}_{mortality} \mid \boldsymbol{\theta}_{mortality}, \boldsymbol{\theta}_{mortalityLag}, \boldsymbol{\theta}_{scren\_effect}\right)$$

$$= \prod_{p,c,r,s} \frac{\left(\mu_{p,c,r,s}\right)^{X_{p,c,r,s}} \cdot e^{-\mu_{p,c,r,s}}}{X_{p,c,r,s}!} \qquad (11)$$

Here the expected number of cases in each combination, $\mu_{p,c,r,s}$, is dependent on both the number of person years at risk and the hazard rate in the given combination of $p$, $c$, $r$ and $s$. Defining $\lambda_{p,c,r,s}$ as the rate of the cause specific deaths for period $p$ among cohort $c$ in region $r$ with a screening history $s$ we now get:

$$\mu_{p,c,r,s} = \lambda_{p,c,r,s} \cdot PY_{p,c,r,s} \qquad (12)$$

Where $PY_{p,c,r,s}$ is the person years at risk for period $p$ among cohort $c$ in region $r$ with screening history $s$.

Borrowing on the formulas from the refined mortality regression analysis of screening effects (Method II), we could apply formula (6), (7) and (8) to find $\mu_{p,c,r,s}$. As the likelihood has special parameters for the time from diagnosis to cause specific death in the absence of screening, we exchange the $\hat{\rho}_{\delta,ai}$ estimator with a function of the corresponding model parameters. Combining the probability of different lags, we define the probability of a lag longer or equal to $\delta$ months from diagnosis to cause specific death in the absence of screening as:

$$\iota_{\delta,a_j} = 1 - \sum_{i=1}^{i=\delta-1} \delta_{i,a_j} \tag{13}$$

where $\delta_{i,a_j}$ is the probably of a lag from diagnosis to cause specific death in the absence of screening of $i$ months, given in $\boldsymbol{\theta}_{mortalityLa} = \left[\delta_{1,a_1}, \delta_{2,a_1}, \delta_{3,a_1}, \ldots, \delta_{k,a_g}\right]$, with $a_j$ as the age group under study.

For the age-cohort-region population data, we now apply suitable $\iota_{\delta,a_j}$'s for each cohort, period and region combination. Defining $L_{c,p,r}$ as $\iota_{\delta,a_j}$ for the age group $a_j$ covering the given age $p - c$, with $\delta$ equalling the observed time from screening invitation for the given cohort $c$ in period $p$ for region $r$. Applying $L_{c,p,r}$, we know get the following formulas for the expected hazard rates:

$$\lambda_{p,c,r,pre} = \exp(\ A(a) + P(p) + C(c) + R_r\ ) \tag{14}$$

$$\lambda_{p,c,r,postold} = \exp(\ A(a) + P(p) + C(c) + R_r\ ) \times L_{c,p,r} \tag{15}$$

$$\lambda_{p,c,r,postnew} = \exp\big(\ A(a) + P(p) + C(c) + R_r + ScrEff \cdot Scr_{c,p,r}\ \big) \tag{16}$$
$$\times \big(1 - L_{c,p,r}\big)$$

Where $pre$ represent pre-screening, $postold$ represent post-screening with pre-screening diagnosis, and $postnew$ represent post-screening with post-screening diagnosis.

**Likelihood for the observed lag from diagnosis to cause specific death:**

Counting the observed lags from diagnosis to cause specific death by months, we can arrange the data on time lags as $\boldsymbol{data}_{mortalityLa} = [NoLag_o, NoLag_1, NoLag_2, \ldots, NoLag_M]$, where $NoLag_i$ is the $M$ frequencies of cases with $i$ months lag between diagnosis and cause specific death pre-screening. Since these counts data represents the number of events distributed across mutually exclusive time intervals, their joint likelihood follows a multinomial distribution given as:

$P\left(\boldsymbol{data}_{\text{mortalityLag}} \mid \boldsymbol{\theta}_{\text{mortalityLag}}\right)$

$$= \frac{\left(\sum_{i=1}^{i=M} NoLag_i\right)!}{\prod_{i=1}^{i=M}(NoLag_i!)} \cdot \left(\prod_{i=1}^{i=(M-1)} \delta_{i,a_j}^{NoLag_i}\right) \qquad (17)$$

where $\delta_i$ represents the probability of a lag of *i* months,

$\boldsymbol{\theta}_{\text{mortalityLag}} = \left[\delta_{1,a_1}, \delta_{2,a_1}, \delta_{3,a_1}, \ldots, \delta_{k,a_g}\right]$,

and $a_j$ as the age group under study.

**Estimating the model parameters using numerical optimization**

Likelihood estimation is widely applied in statistical practice[36,38], and usually numerically effective on modern computers maximizing the log-likelihood. However, implementing the above presented likelihood in the R statistical software we found it difficult to optimize. Using reasonable starting values based on our knowledge of the given data, the log likelihood in practice underflowed to zero preventing numerical optimization. Hence, we used the adjusted regression approach (method II) estimates as starting values in the numerical optimization. As with the refined mortality regression analysis (method II), we deduced confidence intervals using bootstrap replications.

## 6. *Estimating screening effects using Norwegian and (new) Danish data*

### 6.1. *Data for estimating screening effects:*

To compare the different estimation techniques, we applied both the Norwegian data from Weedon-Fekjær et. al. 2014[19] and additional Danish data. For each dataset we estimated the screening effect using both a non-refined mortality analysis and the three here suggested estimation approaches. As both data sources have earlier been analysed using different refined mortality techniques based on selected comparison groups[7,8,17,40], we could also evaluate the efficacy of the newly suggested method in comparison with classic refined mortality analysis approaches.

In Denmark, the public breast cancer screening program started in the municipality of Copenhagen on 1 April 1991[7], inviting women aged 50-69 years every second year. Following this, screening started in the Funen county and Frederiksberg municipality in 1993 and 1994, respectively. Otherwise, no screening program was implemented during the first ten years of the Copenhagen programs[41]. Later Bornholm was added to the screening program in 2001, Vestsjælland in 2004, and the remaining part of Denmark between 2007 and 2010[42]. Combined with little opportunistic screening[43], Denmark hence have good data for evaluating mammography screening.

The study groups in this article include Danish women 30-90 years of age during at least part of the time from 1 April 1981 to 31 March 2001. Once a woman had been included in a screening study group, she remained there even if she moved to another region. We followed up all women from their first date of invitation until death, emigration, or 31 March 2001. Data on underlying cause of death came from the Danish Cause of Death Register. For each woman dying from breast cancer, we identified date of diagnosis with breast cancer by linkage with the Danish Cancer Register.

In Norway, public mammography breast cancer screening was introduced gradually between 1995 and 2005 as part of the BreastScreen Norway program. This study includes all Norwegian women aged 50 to 79 years between 1986 and 2009. For more about the applied Norwegian data, see Weedon-Fekjær et. al. 2014[19].

## 6.2.    *Estimated screening effects:*

The diluted non-refined mortality approach estimated much smaller screening effect than the refined mortality approaches in both Norwegian and Danish data (Table 1). There is some non-systematic variation among the suggested estimators, but only minor differences between the adjusted regression approach (method II) and the full maximum likelihood approach (method III).

Table 1: Comparison of estimated screening effects using different estimating approaches on Norwegian and Danish screening data

| No | Estimation method | Norwegian data | Danish data |
|----|-------------------|----------------|-------------|
| 0 | Expected vs. observed post-screening mortality without splitting breast cancer death by time of diagnosis **(biased, underestimating the screening effect)** | 0.94 | 0. 86 |
| I | Expected vs. observed post screening mortality, using refined mortality | 0.79 | 0. 79[*] |
| II | Regression with refined mortality using offsets based on the expected proportion of cases with diagnosis pre and post first screening invitation | **0.72** | **0.81** |
| III | Full maximum likelihood estimating using refined mortality | 0.72[**] | 0.79 |

[*]) Estimated to 0.75 in Olsen et al. 2005[7] using slightly different data and individual dates on first invitation as basis for calculating persons-years.

[**]) Following Norwegian data privacy regulations, the individual data from the Norwegian Mammography Screening evaluation was deleted at the end of the evaluation. As the maximum likelihood approach needs slightly different input data, estimates is based on recreated approximated data.

Comparing the statistical precision of the here suggested approaches with the earlier refined mortality approaches based on selected comparison groups[7,8,17,40], confidence interval widths are reduced by 46% to 63% using Norwegian data (Table 2). On Danish data, with shorter post screening observation time and fewer counties included in our given observation period, there is a 15% reduction in confidence interval width.

Table 2:    Statistical uncertainty for different estimators of screening effectiveness on reducing breast cancer mortality, all using the recommended refined mortality data. The estimators uncertainty is reported for the gradually introduced Norwegian BreastScreen program, and the Danish program which was mostly rolled out in two steps.

| Study (reference) | Estimation approach (study design) | Width of 95% confidence interval (absolute in percent points) |
|---|---|---|
| **Norway:** | | |
| Kalager et al. 2010 [17] | Selected comparison groups | 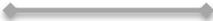 CI* width: 28 % |
| Olsen et al. 2012 [8] | Selected comparison groups | 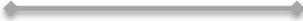 CI* width: 41 % |
| Weedon-Fekjær et al. 2014 [19] | *New adjusted regression approach* | 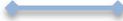 CI* width: 15 % |
| **Denmark:** | | |
| Olsen et al. 2005 [7] | Selected comparison groups | 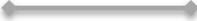 CI* width: 26 % |
| New estimate (current study) | *New adjusted regression approach* | 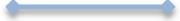 CI* width: 22 % |

* Confidence interval

Applying the different estimation techniques, our experience was that the maximum likelihood approach needed considerable programming and care with the starting values for the optimization. Compared to the full likelihood approach, the adjusted Poisson regression approach was easy to implement in standard statistical software giving almost the same estimated values (Table 3).

Table 3: Comparison of methods for estimating screening effects: Strengths and weaknesses

| No | Estimation method | Strengths | Weaknesses |
|----|-------------------|-----------|------------|
| 0 | Expected vs. observed post-screening rates (using general breast cancer mortality trends) | • Intuitive<br>• Does not require mortality data split by diagnosis pre vs. post screening | • **Highly diluted screen effect estimate** (as estimated effect is not specific to post screening incident cases with a potential screening effect) |
| I | Expected vs. observed post screening, using refined mortality | • Consistent estimate<br>• Intuitive, simple | • Does not fully utilize the available data<br>• Require data on time from diagnosis to death in the absence of screening |
| II | Refined mortality regression with offset based on the expected proportion of cases with diagnosis pre and post first screening invitation **[Recommended]** | • Consistent estimate<br>• Applies all relevant data | • Require data on time from diagnosis to death in the absence of screening |
| III | Full maximum likelihood estimating using refined mortality | • Consistent estimate<br>• Applies all relevant data<br>• Build on standard statistical maximum likelihood theory | • Not easily implemented in standard software<br>• Numerically intensive likelihood optimization<br>• Require data on time from diagnosis to death in the absence of screening |

## 7. Discussion

As discussed in earlier studies[13,25,40], statistical analysis of population screening programs should be analysed using refined mortality to avoid diluted estimates of the true causal screening effect. Historically, refined mortality studies have, however, only managed to apply parts of the available data using selected comparison groups[7,8,17]. Refined mortality studies assume similar distribution of lag time from clinical breast cancer diagnosis to breast cancer death in the absence of screening across comparison groups. We here further apply this assumption to also utilize all the available data from population screening programs (Figure 4). By aligning pre- and post-screening incident cases based on the observed historic lag from cancer diagnosis to death, we here deduce estimates of screening effectiveness on reducing mortality using Poisson regression with specially crafted regression offsets. On Danish and Norwegian data this suggested approach reveals narrower confidence intervals than other refined mortality approaches. This is particularly evident for the gradually introduced Norwegian screening program, laying the ground for improved clinical decision making.

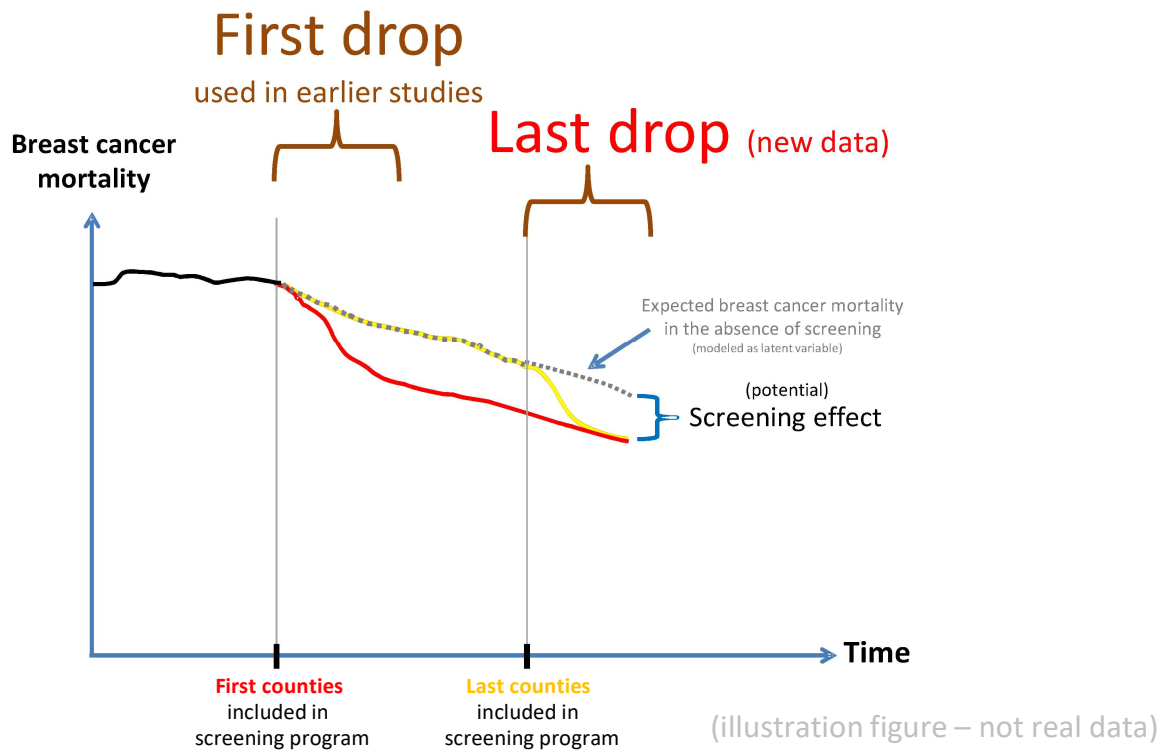

Figure 4:    Illustration of the two expected drops in cancer mortality trends following a two-step implementation of a screening program. Mark that the effects on non-refined mortality are exaggerated to more clearly show the contrasts available for the statistical modelling.

As the earlier applied methods, the new approach does have a set of key modelling assumptions. Most importantly, breast cancer mortality is assumed to follow one overall trend across regions screened early and late. This might be reasonable, but with non-randomized screening introductions there is always a risk that the baseline mortality trend differ by time of screening introduction. Counties with early screening introduction might also be faster in applying new treatments. Ideally, screening introduction should be randomized. One smart way of screening introduction laying the ground for later evaluations, is mixing start up times between different birth cohorts as seen with breast cancer mammography screening in Finland[15].

In our analysis it is assumed that screening modifies the mortality rate of breast cancer for all post-screening incident cases by a proportional multiplicative factor. This might or might not be correct. In practice, the screening effect might vary by time since screening, possibly by postponing some breast cancer death to a later time point. Screening might also use some time to manifest its effect among new incident cases, as the clinical improved prognosis might only be substantial among cases with long lead-time due to detection at screening. On the other side, the relative effect on new cases can be highest soon after screening initiation, as slowly developing cancer with long time from diagnosis to death is mostly low stage cancer at clinical diagnosis with little potential for improved survival by earlier diagnosis. With enough data available, we might model variations in screening effects by time using smoothing splines, but typically the statistical power of most datasets limits this possibility.

With widely available screening methods as mammography, there is also a risk of opportunistic screening outside the official program impacting the estimated screening effect. If opportunistic screening is already well distributed into the population before the official program, the estimated screening effect only becomes an estimate of the additional effect of organized screening. In addition to the potential diluting of the estimated screening effects, opportunistic screening might in some rare cases actually bias the screening effect upward. As pointed out by the Research-based evaluation of the Norwegian Breast Cancer Screening Program[44], opportunistic screening peaking just before introducing organized screening might move cancer from the post-screening evaluation group to the pre-screening comparison group, leading to an overestimated screening effect on the reduction of mortality for post screening incident cases. Generally, unregistered opportunistic screening complicates evaluations of public screening programs and should ideally be reported to public health registries.

While we here focus on screening, the suggested approach might also be adjusted for other settings of delayed intervention effects. Compared to traditional methods for refined mortality analysis of screening programs, the here presented method improves the estimation, but as always with non-randomized epidemiological data care should be taken in the causal interpretation of single studies.

## 8. *Summary; Recommendations*

To secure non-diluted estimates, any population study of screening effects should apply estimates based on refined mortality. Compared to the classical refined mortality methods, the here suggested adjusted regression approach applies more of the available data, resulting in considerable improved precision of the estimated screening effect. We would especially recommend this approach for data from populations with complex screening introduction patterns, where analysis of select comparison groups might discard substantial parts of the available data. While the maximum likelihood approach usually is a good estimator, it is hard to implement for the given refined mortality analysis making the regression approach with offsets our recommended method for analysis of screening program data.

## 9. *Funding*

There was no specific funding for the given theoretical work, but the original non-maximum likelihood approach was developed working at the Norwegian Research Council's "Research-based evaluation of the National Mammography Screening Program" (grant reference No 189503).

## 10. *Declaration of conflicting interests*

The authors declare that there is no conflict of interest.

## 11. *Acknowledgements*

We want to thank now deceased Lars Vatten for good leadership of the Norwegian University of Science and Technology (NTNU) part of the Norwegian Research council's evaluation of Breast Screening Norway, whereby this statistical approach was developed.

In addition, we want to thank Søren Nymand Lophaven for his great work making the Danish data available for our analysis, and Ørnulf Bogan for important input on the presentation of the maximum likelihood part.

# Appendix: Practical implementation of the refined mortality regression analysis of screening effects (Method II) using R and Python

After carefully formatting the data, the screening effect can be estimated using a few lines of code in standard statistical software. First, the observed mortality is summarized with separate lines for period, cohort, region, and screening history (pre-screening mortality, post-screening mortality based on diagnoses pre-screening, and post-screening mortality based on diagnoses post-screening). Then, an indicator for potential screening effect, person years at risk, and our specially crafted offsets to adjust for the differences in expected mortality after splitting cases by screening history are added.

In the freely available R statistical program (http://www.r-project.org), the screening effect can now be estimated by:

```
# Load required libraries:
> library(splines)

# Calculate synthetic age from year and cohort:
> ScrData$Sage <- ScrData$year - ScrData$cohort

# Estimate model using natural smoothing splines and our custom model offsets:
> estScr <- glm(cases ~ ns(year,5) + ns(cohort,5) + ns(Sage,5) + as.factor(region) +
                scrInd + offset(log(personYears)) + offset(log(propTarget)) -1,
                family=poisson(), data=ScrData)

# Extract the estimated relative risk:
> exp(estScr$coefficients[grep("scrInd",names(estScr$coefficients))])
```

Where:

| | |
|---|---|
| cases | is the number of breast cancer deaths (in the given combination) |
| year | is calendar year |
| cohort | is birth cohort |
| region | is region of residence |
| scrInd | is an indicator for potential screening effect, only one for the post-screening mortality that is based on post-screening incident cases |
| personYears | is number of women / person year (in the given combination) |
| propTarget | is for the screening group the estimated probability that a breast cancer death belongs to the given group of cases incident pre- or post-screening assuming that invitation to screening has no effect on breast cancer mortality (estimated based on historic data for each region, cohort and year), and one for the non-screening group. |

Similarly, you might use the common Python programming language (https://www.python.org/):

```python
# Load required libraries:
import pandas as pd
import numpy as np
import statsmodels.api as sm
import statsmodels.formula.api as smf

# Calculate synthetic age from year and cohort:
ScrData["Sage"] = ScrData["year"] - ScrData["cohort"]

# Estimate model using natural smoothing splines and our custom model offsets:
formula = (
    "cases ~ "
    "cr(Sage, df=5) + cr(cohort, df=5) + cr(year, df=5) + C(region) + scrInd - 1"
)
model = smf.glm(
    formula=formula,
    data=ScrData,
    family=sm.families.Poisson(),
    offset=np.log(ScrData["personYears"] + np.log(ScrData["propTarget"])
)
estScr = model.fit()

# Extract the estimated relative risk:
print(np.exp(estScr.params["scrInd"]))
```